\documentclass[showpacs,preprintnumbers,amsmath,amssymb]{revtex4}
\usepackage{graphicx}% Include figure files
\usepackage{bm}% bold math
\newcommand{\beq}{\begin{equation}}
\newcommand{\eeq}{\end{equation}}
\newcommand{\bs}{\boldsymbol}
\newcommand{\eks}{\mathbf{r}}
\newcommand{\dis}{\mathbf{u}}
\newcommand{\rp}{\rm p}
\newcommand{\rs}{\rm s}
\newcommand{\norm}{\hat{\mathbf{n}}}
\newcommand{\tang}{\hat{\mathbf{t}}}
\newcommand{\trac}{\mathbf{t}}
\newcommand{\intc}{\int_{\cal{C}}}
\newcommand{\Nat}{\mbox{I\!\!N}} %natural numbers 

\begin{document}

\title{
Short wave length approximation of a boundary integral operator
for homogeneous and isotropic elastic bodies
}
\author{ Gregor Tanner$^{\dagger}$ and
Niels S\o ndergaard$^{\dagger\dagger}$
}

\affiliation{
$^{\dagger}$School of Mathematical Sciences,
University of Nottingham, Nottingham NG7 2RD, UK,\\
$^{\dagger\dagger}$Div.\ of Mathematical Physics,
Lund Technical University, Sweden.
}
\date{\today}
\begin{abstract}
We derive a short wave length approximation of a boundary integral 
operator for two-dimensional isotropic and homogeneous elastic bodies of 
arbitrary shape. Trace formulae for elastodynamics can be deduced in 
this way from first principles starting directly from Navier-Cauchy's 
equation.
\end{abstract}

\pacs{05.45.Mt,62.30.+d,03.65.Sq,46.40.Cd}

\maketitle
\section{Introduction}
\label{sec:introduction}
We consider the eigenfrequency spectrum of isotropic and homogeneous
elastic bodies of arbitrary shape in two dimensions. The governing
linear equations, the Navier-Cauchy equations, are separable only for
a very small set of geometries such as spherical bodies or infinitely
long cylindrical wave guides. Solutions to the vast majorities of
shapes can be obtained only with the help of numerical techniques such
as finite element or boundary integral methods
\cite{Kit85,Bon95}. Purely numerical approaches are, however, severely
limited by computer resources and often restricted to the low
frequency regime with wavelengths only one or two orders of magnitudes
smaller than the typical size of the system. In the high frequency
limit statistical methods such as {\em statistical energy analysis}
\cite{SEA} or {\em random matrix theory} \cite{Meh91} have proved
valuable. While the former yields information about mean response
signals neglecting interference effects, the latter provide answers
regarding the universal part of the fluctuations in the signal not
taking into account system dependent effects. An alternative approach
providing
more detailed information in the mid to high frequency regime is
obtained using asymptotic methods.

Similar to geometric optics, asymptotic methods connect wave
propagation to an underlying ray dynamics in the small wave 
lengths limit. In elasticity,
typical boundary conditions such as free boundaries lead to mode
coupling and ray splitting at boundaries which complicate a ray
analysis. The bulk of the asymptotic analysis in the
elastodynamics literature has focused on interface or scattering
problems \cite{Ach82} in which the number of reflections for a typical
path is small. A ray-treatment of interior modes of elastic bodies of
finite size has received much less attention so far; here, geometric
rays have an infinite number of reflections with the boundaries
undergoing ray-splitting at every impact which leads to summation and
ordering problems when expressing operators such as the Green function
in a short wavelength approximation.

Such issues have been addressed in the context of the Helmholtz
equation in finite domains and more generally in quantum mechanics. Especially,
the connection between the solution of these scalar wave equations
and the dynamical properties of a related ray or classical dynamics
have been treated in much detail in the context of {\em quantum chaos}
\cite{Gut90,Stoe99,TRR00}. A powerful tool connecting the spectrum of a
quantum system with an underlying classical dynamics are trace
formulae as for example introduced by Gutzwiller in quantum mechanics 
\cite{Gut90}, expressing the trace of the Green function in terms of the 
periodic orbits of the classical system.

A trace formula for the interior problem in elasticity has been
presented first in \cite{Cou92}; the result was, however, obtained by
way of comparison with the scalar Helmholtz equation and not derived
from the governing equations.  To the best
of our knowledge, such a derivation is still lacking, which is
desirable not only from a point of principle, but is essential to
obtain corrections beyond the leading large wavenumber asymptotics. A
powerful technique for deriving such trace formulae from the
underlying wave equations is to derive asymptotic expressions for
boundary integral operators in terms of semiclassical 
{\em transfer operators}, a method pioneered by Bogomolny
\cite{Bog92}, see also \cite{Dor92,Pro94,Rou95}.  The general form of
such a transfer operator for elastic problems has been postulated in
\cite{ST02} and verified for the special case of a circular
waveguide. A derivation of the transfer operator for the biharmonic
equation describing the out-of-plane vibrations of plates has been
obtained in \cite{BH98} incorporating the coupling of flexural and 
boundary modes. In the
short wave length limit, the wave equation reduces again to a
scalar problem with modified boundary conditions due to the
exponential damping of surface waves away from the boundary.

In the following we derive the Bogomolny transfer operator and form 
there periodic orbit trace formulae for two dimensional elasticity starting 
from first principles.

\section{The transfer operator}
\label{sec:transfer}
\subsection{Fundamental equations}
We  consider  isotropic and homogeneous elastic
bodies described in the frequency domain by
the Navier-Cauchy equation \cite{LL59}
\beq
\mu \Delta \dis + (\lambda + \mu) {\bf \nabla ( \nabla \cdot \dis)} +
\rho \omega^2 \dis = 0 \,,
\label{pde}
\eeq
where $ \dis(\eks)$ is the displacement field, $\lambda$, $\mu$ are the
material dependent Lam\'e coefficients and $\rho$ is the density which we
assume to a constant. We will consider free boundary
conditions here, that is no forces act normal to the boundary; this can be
expressed in terms of the traction $\trac(\dis)$, that is,
\beq \label{BC}
\trac(\dis)= \norm \cdot \bs{\sigma}(\dis) = 0
\eeq
where $\norm$ is the normal at $\eks$ on the boundary 
$\cal{C}$ of the elastic body; the stress tensor $\bs{\sigma}(\dis)$ is given as
\beq
\bs{\sigma}(\dis)= \lambda (\nabla \cdot \dis) \, {\mathbf 1} +
\mu(\nabla \otimes \dis + \dis \otimes \nabla) \, .
\eeq
We make the standard Helmholtz decomposition of the displacement field $\dis$,
that is,
\beq \label{helm}
{\bf u} = {\bf u}_{\rp} + {\bf u}_{\rs}
\quad \mbox{with} \quad
{\bf u}_{\rp} = \nabla \Phi, \;\;
{\bf u}_{\rs} =  {\bf \nabla} \times {\bs{\Psi}} \; ;
\eeq
the elastic potentials $\Phi$ for the pressure (or longitudinal) and
${\bs{\Psi}}$ for the shear (or transversal) wave component solve
Helmholtz's equation 
\beq
 \begin{array}{c} (\Delta+k_{\rp}^2) \Phi =0 \\ 
                  (\Delta+k_{\rs}^2) \Psi = 0 
  \end{array} 
\eeq
with wave numbers $k_{\rp}$ and $k_{\rs}$, respectively.
One finds the dispersion relation $k_{\rp,\rs} = \omega/c_{\rp,\rs}$ with wave
velocities
\beq \label{velo}
c_{\rp}=\sqrt{\frac{\lambda + 2 \mu}{\rho}} \hspace{1cm}
c_{\rs}=\sqrt{\frac{\mu}{\rho}} .
\eeq

In the following, we shall restrict ourselves to two-dimensional
problems, that is $\eks, \dis(\eks) \in \mathbb{R}^2$ and we set
${\bs{\Psi}} = (0,0,\Psi)^t$. The resulting differential equations
describe in-plane deformations in plates or wave propagation in bodies with
fixed shape in the $xy$ plane extending to $\pm \infty$ along $z$. 

\subsection{Boundary integral equations}
\subsubsection{General set-up}
In what follows, we will adapt the method outlined in \cite{Bog92, BH98}
to the Navier-Cauchy Eqn.\ (\ref{pde}). We first rewrite
the boundary conditions (\ref{BC}) in terms of boundary integral equations
and then consider the Fourier coefficients of the boundary integral
functions.  We start by introducing the elastic potentials in the form
\begin{eqnarray}\label{phi}
\Phi(\eks) &=& \intc G(\eks,\alpha;k_{\rp}) g(\alpha)d\alpha;\\
\label{psi}
\Psi(\eks) &=& \intc G(\eks,\alpha;k_{\rs}) h(\alpha) d\alpha \, .
\end{eqnarray}
where $g$ and $h$ are yet unknown single layer distributions on the
boundary and $\alpha \in [0,L_C]$ parameterises the boundary of length
$L_C$, that is, $\eks(\alpha) \in \cal{C}$; furthermore,
$G(\eks,\eks';k)$ is a Green function solving the inhomogeneous Helmholtz equation
\[ (\Delta +k^2) \, G(\eks,\eks';k) = \delta(\eks - \eks')\, .\]
The integrals converge for $\eks$ inside $\cal{C}$ and non-singular layer
distributions $g$ and $h$, and the ansatz (\ref{phi}), (\ref{psi}) thus
solves the Helmholtz equation in the interior.
A convenient choice for $G(\eks,\eks';k)$ is
the free Green function which in 2 dimensions takes the form
\beq
G(\eks,\eks',k) =\frac{1}{4 i}\, H^{(1)}_0(k|\eks-\eks'|) \, 
\eeq
where $H^{(1)}_0$ is the 0-th order Hankel function.

\begin{figure}[h]
   \centering
   \includegraphics[scale=0.6]{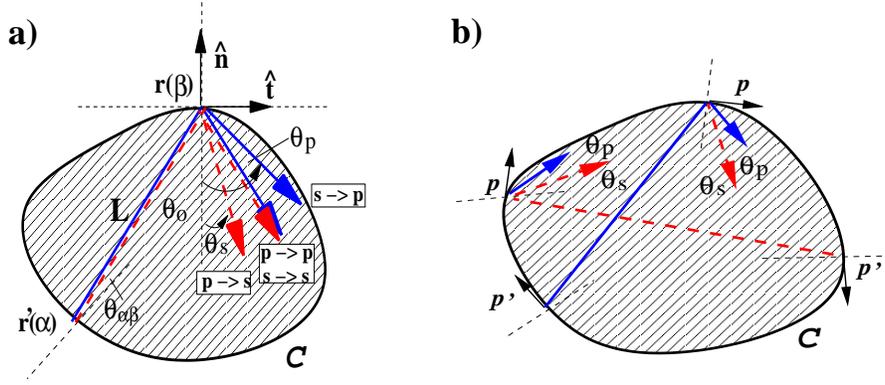} % requires the graphicx package
   \caption{Coordinates on the boundary: a) position representation
    with path of length $L$ from $\eks'(\alpha)$ to $\eks(\beta)$ (here for a
    initial pressure wave); b) momentum representation with shear and pressure 
    path starting with tangential momentum $p'$ and ending with momentum 
    $p$ on the boundary.}
   \label{fig:coord}
\end{figure}

In a next step, it is useful to rewrite the boundary condition (\ref{BC})
in terms of the elastic potentials. Defining $\norm$ and $\tang$ as
the (outward) normal and tangent vectors at the
boundary point $\eks(\beta) \in \cal{C}$ as indicated in Fig.\ 
\ref{fig:coord}a,
one obtains
\beq \label{BCpot}
\norm \cdot \bs{\sigma} = -\lambda\, \norm \,k_{\rp}^2 \Phi +
2 \mu \left[ \norm\,\frac{\partial^2}{\partial n^2} \Phi +
\tang\, \frac{\partial^2}{\partial n\partial t} \Phi +
\norm\,\frac{\partial^2}{\partial n\partial t} \Psi +
\tang\,\frac{1}{2}\left(\frac{\partial^2}{\partial t^2} -
\frac{\partial^2}{\partial n^2}\right)\Psi\right] = 0\, ,
\eeq
where we used $\Delta \Phi = - k_{\rp}^2 \Phi$ valid in the interior; note that all partial
derivatives are understood as being taken in the interior (after a suitable continuation of
the local coordinates system into the interior) and then taking the limit $\eks \to \eks(\beta)\in C$.

We thus need to determine derivatives of the form
\beq \label{parG}
\partial_{nn}
\intc G(\beta, \alpha) f(\alpha)\, d\alpha;\quad
\partial_{n t}
\intc G(\beta, \alpha) f(\alpha)\, d\alpha;\quad
\partial_{tt}
\intc G(\beta,\alpha) f(\alpha) \,d\alpha\, 
\eeq
with $G(\beta,\alpha) \equiv G(\eks(\beta),\eks'(\alpha),k)$,
$f$ stands for $g$ or $h$, respectively, and the derivatives
are always taken with respect to the first variable $\eks(\beta)$ from
the interior. Note that taking the limit $\eks \to \eks(\beta)$ and
differentiating are non-commuting operations due to the logarithmic
singularity of the Green function for $\beta \to \alpha$.

For a short wavelength analysis, we distinguish between long
segments with $k |\eks(\beta) - \eks'(\alpha)| \gg 1$ and
short contributions with $k |\eks(\beta) - \eks'(\alpha)| = {\cal O}(1)$;
for the former, one can employ the asymptotic form of the Green function
\beq \label{Gasym}
G(\eks,\eks',k) \sim
\frac{1}{4 i}\, \sqrt{\frac{2}{\pi k|\eks-\eks'|}} e^{i(k|\eks-\eks'|-\pi/4)}
\qquad k|\eks-\eks'| \to \infty\, ,
\eeq
whereas the logarithmic singularity $G(\eks,\eks',k) \sim \frac{1}{2\pi}
\ln(k|\eks-\eks'|)$ for $k|\eks-\eks'| \to 0$ calls for a separate
treatment for short length contributions.  We note in particular, that one
obtains from (\ref{Gasym}) in leading order,
\beq \label{long}
\partial_n G(\beta, \alpha) \sim i q \, G(\beta, \alpha);\quad
\partial_t G(\beta, \alpha) \sim i p \, G(\beta, \alpha),
\eeq
and likewise for the second order derivatives. Here, $q(\beta,\alpha) =
k \cos\theta_0$ and $p(\beta,\alpha) = k \sin\theta_0$ are the normal and
tangential component of the wave vector ${\mathbf k} = k \,(\eks(\beta)
- \eks'(\alpha))/ |\eks(\beta) - \eks'(\alpha)|$ at the boundary point
$\beta$, see Fig.\ \ref{fig:coord}a. 

\subsubsection{Asymptotic form of the boundary integral kernel in
momentum representation}
Following \cite{BH98}, we split the boundary integral into two parts, that is,
\[ \intc d\alpha = \int_{{\cal C}/\Delta}d\alpha + \int_{\Delta}d\alpha\]
where $\Delta$ refers to a small interval around $\alpha = \beta$ scaling as
$\Delta \sim k^{-1 + \epsilon}$ with $0< \epsilon<1$.

We deal with the short length contributions first. Due to the scaling chosen
for the interval $\Delta$, we can neglect curvature contributions in the large
$k$ limit and write in leading order in $1/k$
\beq  \label{short}
\int_{\Delta} G(\eks(\beta),\eks(\alpha),k) f(\alpha) d\alpha \sim
-\frac{1}{k}\int^{k\Delta/2}_{-k\Delta/2} G(0,x/k,k) f(x/k) dx \sim
-\frac{1}{k} \int_{-\infty}^{\infty} G(0,x/k,k) f(x/k) dx
\eeq
thus integrating along a straight line in direction of $\tang(\beta)$
centred at $\eks(\beta)$. It is now convenient to express the free
Green function in integral representation, which in two dimensions leads to
\beq
G(\eks,\eks',k) = - \lim_{\epsilon\to 0} \int \frac{dp^2}{4 \pi^2}
\frac{e^{i{\mathbf p} \cdot (\eks - \eks')}}{p^2 -(k^2 +i\epsilon)} \; .
\eeq
Aligning the $x$-axis with the tangential direction $\tang(\beta)$ as in
(\ref{short}) and integrating out the $p_y$ component, one obtains
\beq\label{1d}
G(\eks,\eks',k) = \int \frac{dp}{2 \pi} e^{i p (x - x')}
\frac{e^{i q |y - y'|}}{2 i q}
\eeq
with $q = \sqrt{k^2 - p^2}$. Note that $\lim_{y\to 0_-}\partial_y G(\eks,\eks')
= \frac{1}{2} \delta(x-x')$ revealing the singular behaviour of the Green function
in this limit.

Next, we express the single layer distributions on the boundary in its
Fourier components, that is,
\begin{equation}
f(\alpha) = \int dp \, \hat{f}_p e^{i p \alpha}
\end{equation}
where we treat $p$ to leading order as a continuous variable neglecting 
the discreteness of $p = 2\pi j/L_C, j\in \Nat$ due to the finite length of 
the boundary $\cal{C}$. From (\ref{1d}) together with (\ref{short}), we 
obtain the short length contributions in the form
\begin{equation}\label{shortf}
\int_{\Delta} G(\beta,\alpha) f(\alpha) d\alpha \sim \frac{1}{k}\lim_{y\to 0_-}
\int dp dp' \, e^{i p \beta}\frac{e^{i q |y|/k}}{2 i q}
\int_{-\infty}^{\infty}\frac{dx}{2\pi}\, e^{i (p-p') x/k} \hat{f}_{p'}
= \int dp \frac{\hat{f}_{p}}{2 i q} e^{i p \beta}\, .
\end{equation}
We proceed as above for the partial derivatives (\ref{parG}) by identifying the normal
and tangential direction with the $y$ and $x$ axis, respectively.  Note again, that the
derivatives $\partial/\partial y$ need to be taken before completing the limit $y \to 0_-$ from
below.  One obtains
\begin{eqnarray}
\partial_{nn}
\int_{\Delta} G(\beta,\alpha) f(\alpha) d\alpha &=& i \int dp\, \hat{f}_p \frac{q}{2}e^{i p \beta};\\
\partial_{nt}
\int_{\Delta}G(\beta,\alpha) f(\alpha) d\alpha &=& - i \int dp\, \hat{f}_p\frac{p}{2}e^{i p \beta};\\
\partial_{tt}
\int_{\Delta} G(\beta,\alpha) f(\alpha) d\alpha &=& i \int dp\, \hat{f}_p\frac{p^2}{2 q}e^{i p \beta}.
\end{eqnarray}

Turning to the contributions from long trajectories, we again introduce the 
Green function on the boundary in terms of its Fourier components
\begin{equation}
G(\beta,\alpha,k_{\rp/\rs}) = \int dp dp' \, \hat{G}^{\rp/\rs}_{pp'}
e^{i(p\beta  - p' \alpha)} 
\end{equation}
and write
\[
\int_{{\cal C}/\Delta}d\alpha \,G(\beta, \alpha) f(\alpha) = \int dp dp'
\hat{G}_{pp'} \hat{f}_{p'} e^{i p \beta}\, .
\]
Here, differentiation can be pulled under the integral sign and by employing 
the asymptotic form (\ref{Gasym}) together with a stationary phase 
approximation, one obtains in leading order
\[ \widehat{\partial_{n} G}_{pp'} = i p \hat{G}_{pp'};\quad \widehat{\partial_{t} G}_{pp'}
= i q \hat{G}_{pp'}\]
and likewise
\[
\widehat{\partial_{nn} G}_{pp'} = - p^2 \widehat{G}_{pp'};\quad
\widehat{\partial_{t n} G}_{pp'} = - q p \widehat{G}_{pp'}; \quad
\widehat{\partial_{tt} G}_{pp'} = - q^2 \widehat{G}_{pp'}\, .
\]
Writing the boundary conditions (\ref{BC}) in terms of the Fourier components 
$\hat{g}_p$, $\hat{f}_p$ and $G^{\rp/\rs}_{pp'}$ for both short and long 
contributions, one obtains the set of equations
\beq
\left({\mathbf M}_0 + {\mathbf M}\hat{\mathbf D}\right) \hat{\mathbf X} = 0 
\qquad \mbox{with} \quad
\hat{\mathbf X}_p = 
\left( \begin{array}{c} \hat{g}_p\\\hat{h}_p\end{array}\right)
\eeq
where
\beq
\left({\mathbf M}_0\right)_{pp'} = \frac{i}{2}\left(\begin{array}{cc}
\frac{1}{q_{\rp}}(\lambda k^2_{\rp} + 2 \mu q^2_{\rp})& -2 \mu p \\
-2 \mu p & \frac{\mu}{q_{\rs}}(p^2 - q^2_{\rs})
\end{array}\right) \delta_{pp'}\, ; \quad
{\mathbf M}_{pp'} =\frac{i}{2} \left(\begin{array}{cc}
\frac{1}{q_{\rp}}(\lambda k^2_{\rp} + 2 \mu q^2_{\rp})& 2 \mu p \\
2 \mu p & \frac{\mu}{q_{\rs}}(p^2 - q^2_{\rs})
\end{array}\right)\delta_{pp'}
\eeq
and
\beq
\hat{\mathbf D}_{pp'} = 2 i \left(\begin{array}{cc}
                        q_{\rp}\,\hat{G}^{\rp}_{pp'}& 0\\
                        0& q_{\rs} \,\hat{G}^{\rs}_{pp'}
                        \end{array}\right)
                      = 2 \left(\begin{array}{cc}
                        \widehat{\partial_n G}^{\rp}_{pp'}& 0\\
                        0& \widehat{\partial_n G}^{\rs}_{pp'}
                        \end{array}\right)\, .
\eeq
The eigenfrequency condition for finite elastic bodies in 2 dimensions can 
thus be cast into the form
\beq \label{det}
\det(\mathbf{I} - \hat{\mathbf{T}}(\omega)) = 0 \qquad \mbox{with} \quad
\hat{\mathbf{T}} = - {\mathbf M}^{-1}_0{\mathbf M}\hat{\mathbf D}\,  \eeq
and
\beq\label{Top}
\hat{\mathbf{T}}_{pp'} =  \frac{1}{4 \det ({\mathbf M}_0)}
\left(\begin{array}{cc}
\frac{\mu}{q_{\rs}q_{\rp}}(p^2 - q_{\rs}^2)(\lambda k_{\rp}^2 + 2 \mu q_{\rp}^2) + 4 \mu^2 p^2&
\frac{4 \mu^2 p}{q_{\rs}}(p^2 - q_{\rs}^2)\\
\frac{4 \mu p}{q_{\rp}}(\lambda k_{\rp}^2 + 2 \mu q_{\rp}^2) &
\frac{\mu}{q_{\rs}q_{\rp}}(p^2 - q_{\rs}^2)(\lambda k_{\rp}^2 + 2 \mu q_{\rp}^2) + 4 \mu^2 p^2
\end{array}\right)
\cdot \hat{\mathbf D}_{pp'}
\eeq
as well as
\beq
\det ({\mathbf M}_0) = -\frac{1}{4}\left[
\frac{\mu}{q_{\rs}q_{\rp}}(p^2 - q_{\rs}^2)(\lambda k_{\rp}^2 + 2 \mu q_{\rp}^2) 
- 4\mu^2 p^2\right] \, .
\eeq
The operator $\hat{\mathbf T}$ is the short wavelength approximation of a wave 
propagator acting on boundary functions in Fourier or momentum representation; 
it has the general form of a quantum Poincar\'e map \cite{Bog92, Pro94}, here
written for the elastodynamic case including mode conversion. The matrix elements 
$\hat{\mathbf T}_{pp'}$ describe the evolution of pressure and shear waves 
along 'ray' - trajectories starting on the boundary with tangential 
momentum $p'$ and hitting the boundary with tangential momentum $p$;
note that the rays corresponding to two different modes will in general
start and end at different points on the boundary,  see Fig.\ \ref{fig:coord}b.
The $q_{\rp/\rs}$ component is the part of the wave vector $\bf k_{\rp/\rs}$ 
normal to the interface and we may set
\[ 
p_{\rp/\rs} = k_{\rp/\rs} \sin \theta_{\rp/\rs}; 
\quad  q_{\rp/\rs} = k_{\rp/\rs} \cos\theta_{\rp/\rs}\, .
\]
The tangential momentum $p$ at the end points is the same for both 
polarisations before and after impact with the boundary and we 
obtain directly Snell's law
\begin{equation} \label{Snell}
p = p_{\rp} = k_{\rp} \sin\theta_{\rp} = 
k_{\rs} \sin\theta_{\rs} = p_{\rs} \, .
\end{equation}
Using  
$\kappa = k_{\rs}/k_{\rp} = c_{\rp}/c_{\rs}$ and identities like
\[
\lambda k_{\rp}^2 + 2 \mu q_{\rp}^2 = (\lambda + 2 \mu) k^2_{\rp} 
\cos 2\theta_{\rs}; \quad p^2 - q_{\rs}^2 = - k_{\rs}^2 \cos 2 \theta_{\rs}\, ,
\]
we may write the pre-factor matrix in the form
\beq \label{A}
{\mathbf A} = - {\mathbf M}^{-1}_0 {\mathbf M} = \left(\begin{array}{cc}
                A_{\rp\rp}& A_{\rp\rs}\\
                A_{\rs\rp}& A_{\rs\rs}
                   \end{array}\right), 
\eeq
with 
\begin{eqnarray} \label{pp}
A_{\rp\rp} = A_{\rs\rs} &=& 
\frac{\sin 2 \theta_{\rs} \sin 2 \theta_{\rp} - \kappa^2 \cos^2 2\theta_{\rs}}
{\sin 2 \theta_{\rs} \sin 2 \theta_{\rp} + \kappa^2 \cos^2 2\theta_{\rs}}\\
\nonumber
A_{\rs\rp} &=&  \kappa^2 \frac{2 \sin 2 \theta_{\rs}\cos 2\theta_{\rs}}
  {\sin 2 \theta_{\rs} \sin 2 \theta_{\rp} + \kappa^2 \cos^2 2\theta_{\rs}}\\
\nonumber
A_{\rp\rs} &=& -\frac{2 \sin 2 \theta_{\rp}\cos 2\theta_{\rs}}
 {\sin 2 \theta_{\rs} \sin 2 \theta_{\rp} + \kappa^2 \cos^2 2\theta_{\rs}}\, .
\end{eqnarray}
The matrix elements of $\mathbf A$ are up to a similarity transformation
equivalent to the standard conversion factors for plane shear or pressure 
waves at impact with a plain interface and free boundary conditions 
\cite{LL59}. Note that we follow here the convention used throughout the 
paper; for example, $A_{\rs\rp}$ denotes the conversion amplitude 
between an incoming $\rp$ - wave and an outgoing $\rs$ - wave.  

Next, we express the transition matrix $\mathbf A$ in a slightly different form
using the transformation
\begin{equation} \label{a_mat}
{\mathbf a} = {\mathbf K}^{-1} {\mathbf A} {\mathbf K}
\quad \mbox{with} \quad
                   {\mathbf K} = \left(\begin{array}{cc}
                     (q_{\rp}/q_{\rs})^{1/4}& 0\\
                    0&  (q_{\rs}/q_{\rp})^{1/4}
                    \end{array}\right)\,
\end{equation}
which leads to a unitary matrix $\mathbf a$. The relations
$a_{\rp\rp}^2 + a_{\rs\rp}^2 = 1 = a_{\rp\rs}^2 + a_{\rs\rs}^2$ 
reflect conservation of wave energy {\it normal to the surface} 
in the presence of mode conversion \cite{LL59}. 

\subsubsection{Asymptotic form of the boundary integral kernel in position 
representation}
It is often convenient to work with the boundary integral kernel in position 
representation; the inverse Fourier transformation of the 
operator $\hat{\mathbf T}_{pp'} = \mathbf{A}_p \hat{\mathbf D}_{pp'}$ again 
taken in stationary phase approximation and employing the asymptotic form of 
the free Green function (\ref{Gasym}), yields
\beq \label{Tpos}
{\mathbf T}(\beta,\alpha) = \frac{1}{\sqrt{2 \pi i L}} 
\cos \theta_0 \;
{\mathbf A}(\beta,\alpha) 
\left(\begin{array}{cc}
                        \sqrt{k_{\rp}}\, e^{ik_{\rp}L}& 0\\
                        0& \sqrt{k_{\rs}} \, e^{ik_{\rs} L}
                        \end{array}\right)\, .
\eeq 
The stationary phase condition picks out contributions from 
shear and pressure waves travelling from $\alpha$ to $\beta$ along
rays of length $L$ intersecting the boundary at $\beta$ with a common
angle $\theta_0$, see Fig.\ \ref{fig:coord}a. In contrast to the
momentum - representation considered earlier, rays leaving the end
point $\beta$ can do so along three different directions with angles
$\theta_0$, $\theta_{\rp}$ and $\theta_{\rs}$. A $\rp$ - polarised wave, for
example, may emerge from $\beta$ at an angle $\theta_{0}$ or
$\theta_{\rp}$ depending on whether the corresponding incoming wave
was a $\rp$ or $\rs$ wave. We thus set $\theta_{\rp} \equiv \theta_0$ in
$A_{\rp\rp}$ and $A_{\rs\rp}$ and $\theta_{\rs} \equiv \theta_0$ in
$A_{\rp\rs}$ and $A_{\rs\rs}$ in Eqn.\ (\ref{pp}) with $\theta_{\rp},
\theta_{\rs}$ given by Snell's law (\ref{Snell}); note that this
implies for example that $A_{\rp\rp} \ne A_{\rs\rs}$ in
general. Rewriting the operator (\ref{Tpos}) in terms of the (now in
general non-unitary) transition matrix $\mathbf a$, one obtains 
\beq
\label{Tpos1} {\mathbf T}(\beta,\alpha) = \frac{1}{\sqrt{2 \pi i L}}
\left(\begin{array}{cc} \cos\theta_0 \; a_{\rp\rp}& \sqrt{\cos\theta_0
\cos\theta_p/\kappa}\; a_{\rp\rs}\\ \sqrt{\kappa\cos\theta_0
\cos\theta_s} \; a_{\rs\rp} & \cos\theta_0 \; a_{\rs\rs}
                        \end{array}\right)\, \cdot \,
                        \left(\begin{array}{cc}
                        \sqrt{k_{\rp}}\, e^{ik_{\rp}L}& 0\\
                        0& \sqrt{k_{\rs}} \, e^{ik_{\rs} L}
                        \end{array}\right)\, 
\eeq
with 
\begin{eqnarray} \label{app}
a_{\rp\rp}  = 
\frac{\sin 2 \theta_{\rs} \sin 2 \theta_{0} - \kappa^2 \cos^2 2\theta_{\rs}}
{\sin 2 \theta_{\rs} \sin 2 \theta_{0} + \kappa^2 \cos^2 2\theta_{\rs}} &;&\quad
a_{\rp\rs} = - \kappa \frac{2 \sqrt{\sin 2 \theta_{0}\sin 2 \theta_{\rp}} \cos 2\theta_{0}}
 {\sin 2 \theta_{0} \sin 2 \theta_{\rp} + \kappa^2 \cos^2 2\theta_{0}}\\
\nonumber
a_{\rs\rp} =  \kappa \frac{2 \sqrt{\sin 2 \theta_{0}\sin 2 \theta_{\rs}} \cos 2\theta_{\rs}}
  {\sin 2 \theta_{\rs} \sin 2 \theta_{0} + \kappa^2 \cos^2 2\theta_{\rs}}
&;&\quad
a_{\rs\rs} =
\frac{\sin 2 \theta_{0} \sin 2 \theta_{\rp} - \kappa^2 \cos^2 2\theta_{0}}
{\sin 2 \theta_{0} \sin 2 \theta_{\rp} + \kappa^2 \cos^2 2\theta_{0}} \, .
\end{eqnarray}

For hyperbolic shapes, that is, for boundaries only admitting isolated periodic 
geometric rays (including mode conversion at the boundary), standard arguments
lead to a description of the traces of the operator $\mathbf T$ in terms of
periodic ray trajectories \cite{Bog92}. One obtains
\begin{equation} \label{trT}
 \mbox{Tr} {\mathbf T}^n = \sum_j^{(n)} {\cal A}_j e^{i S_j - i \mu_j \pi/2}
\end{equation}
where the sum is over all periodic ray trajectories having $n$ reflections at the
boundary with position and polarisations $[(\alpha^j_1,l^j_1), \ldots,
(\alpha^j_n,l^j_n)]$ where $l^j_i = \rp$ or $\rs$ is the polarisation of the
$i$th segment of the periodic ray $j$ leaving the boundary at the point
$\alpha_i$, $i=1,\ldots,n$. Furthermore, one has 
\begin{equation}
S_j =\sum_{i=1}^{n} k_{l^j_i} L^j_i; \quad
{\cal A}_j = {\cal A}_j^{geo} \,\prod_{i=1}^n a_{l^j_{i+1}l^j_{i}}
\end{equation}
taken along a periodic orbit; here $S_j$ is the action of classical mechanics and 
the amplitude ${\cal A}_j$ separates 
into a geometric part ${\cal A}^{geo}_j$ containing information about the spreading of nearby 
trajectories and a mode conversion loss factor.  The traces ${\mathbf T}^n$ contain all 
the information about the spectrum and may be used to construct the density of states or  
express the spectral determinant (\ref{det}).

The operator (\ref{Tpos1}0 can be written in a form more familiar from 
semiclassical 
quantum mechanics. We note that the cosine terms in the amplitudes relate to ray
angles before and after hitting the boundary at $\beta$; each contribution to the 
periodic orbit formula (\ref{trT}) thus contain products of cosine terms along the 
periodic orbit. Following an argument by Boasman \cite{Boa94} developed in the scalar
case, we consider 
\begin{equation} \label{amplitude}
\sqrt{\left|\frac{\partial^2 L(\beta,\alpha)}
{\partial \alpha \partial \beta}\right|} = 
\sqrt{\frac{\cos\theta_{\alpha\beta}\cos\theta_{\beta\alpha}}{L}} 
\end{equation}
with angles $\theta_{\beta\alpha} = \theta_0$ taken at $\beta$ and 
$\theta_{\alpha\beta}$ taken at 
$\alpha$, respectively, (see Fig.\ \ref{fig:coord}a). The traces of the 
operators $\mathbf T$ as in (\ref{Tpos1}) and $\tilde{\mathbf T}$ defined as
\begin{eqnarray} \label{Tpos2}
\tilde{\mathbf T}(\beta,\alpha) &= \frac{1}{\sqrt{2 \pi i}}  \,
\sqrt{\left|\frac{\partial^2 L_{\beta \alpha}}{\partial \alpha \partial \beta}
\right|}\, \mathbf{a}(\beta, \alpha) \cdot
\left(
\begin{matrix}
\sqrt{k_{\rp}} \, e^{i k_{\rp} L_{\beta \alpha}} & 0 \\
0 &  \sqrt{k_{\rs}} \, e^{i k_{\rs} L_{\beta \alpha}} \\
\end{matrix}
\right)   \nonumber \\
\end{eqnarray}
are thus equivalent to leading order. That is, when writing the traces as sum over 
periodic rays as in (\ref{trT}), the cosine terms coincide after multiplication along a 
periodic orbit. Similarly, the extra $\kappa^{\pm1/2}$ terms in the off-diagonal terms
in (\ref{Tpos1}) cancel after one period. This confirms the form of the operator as 
postulated in \cite{ST02} from which the trace formula suggested by Couchman {\em et al}
\cite{Cou92} can be derived by standard means as indicated earlier.

\section{Conclusion}
\label{sec:concl}
We have derived an asymptotic form of the boundary integral kernel in 2d 
elastodynamics from which periodic orbit trace formulae can be deduced using 
stationary phase arguments. It is expected that a 3d version of the asymptotic 
operator can be written in the form (\ref{Tpos2}) using local 
coordinates where the tangential direction lies in the plane spanned 
by the vector $\eks - \eks'$ and the normal at the boundary point $\eks$. 
In deriving the 3d version of the operator (\ref{Tpos2}) one is naturally lead 
to a momentum representation in terms of spherical coordinates; the technical  
difficulties are not expected to exceed these of the 3d quantum case as 
discussed in \cite{Bog92} and \cite{Pro94,Rou95}.  \\[.5cm]
Acknowledgements:\\
We would like to thank the EPRSC for financial support.\\[.5cm]

\end{document}